\documentclass[aps,prl,twocolumn,showpacs,superscriptaddress,groupedaddress]{revtex4}

\usepackage{amsmath,amssymb,amsfonts,subfigure,braket,color,dsfont}
\usepackage[english]{babel}
\makeatletter\AtBeginDocument{\let\@elt\relax}\makeatother 
\usepackage{graphicx} 
\usepackage{sistyle}
\usepackage{listings}
\usepackage{gensymb} 
\usepackage{xcolor}

\begin{document} 
 
\title{Homogeneous Linewidth Behaviour of Narrow Optical Emitters at Sub-kelvin Temperatures}
\author{X. Lin} 
\affiliation{LNE-SYRTE, Observatoire de Paris, Universit\' e PSL, CNRS, Sorbonne Universit\' e, Paris, France}
\author{M. T. Hartman} 
\affiliation{LNE-SYRTE, Observatoire de Paris, Universit\' e PSL, CNRS, Sorbonne Universit\' e, Paris, France}
\author{P. Goldner} 
\affiliation{Chimie ParisTech, Universit\' e PSL, CNRS, Institut de Recherche de Chimie Paris, 75005 Paris, France} 
\author{B. Fang} 
\affiliation{LNE-SYRTE, Observatoire de Paris, Universit\' e PSL, CNRS, Sorbonne Universit\' e, Paris, France}
\author{Y. Le Coq}
\affiliation{LNE-SYRTE, Observatoire de Paris, Universit\' e PSL, CNRS, Sorbonne Universit\' e, Paris, France}
\author{S. Seidelin}\email{signe.seidelin@neel.cnrs.fr}
\affiliation{Univ. Grenoble Alpes, CNRS, Grenoble INP and Institut N\' eel, 38000 Grenoble, France}
\date{\today}

\begin{abstract}

We explore the properties of ultra-narrow spectral holes in ensembles of solid-state emitters in crystals over a range of sub-kelvin temperatures, with a focus on their potential application in frequency stabilization schemes as an alternative to ultrastable cavities. We investigate how the parameters used to burn the spectral hole impact its shape, and how these factors determine the minimum achievable linewidth. In addition to the stability of the hole's center frequency, the linewidth and contrast play a crucial role in frequency locking. At sub-kelvin temperatures, the temperature-dependent $T^7$ broadening from two-phonon Raman scattering is expected to be negligible, and the spectral hole's linewidth should therefore remain constant in this interval. We observe however a linear broadening with increasing temperature, highlighting the need for further investigation into the mechanisms governing the linewidth at ultra-low temperatures.

\end{abstract}


\maketitle


Highly coherent optical solid-state emitters play an essential role in scalable quantum technology. Among these, rare-earth ions doped into crystal matrices have long been recognized for their exceptional optical coherence properties~\cite{Yano1991,Equall1994}, positioning them as key components in a wide array of novel applications. Recent advancements have highlighted their potential as quantum memories \cite{Zhong2017,Laplane2017,Chao2020} and quantum interfaces for photons \cite{Rakonjac2021,Rivera2023}, as well as their integration with other qubit systems, such as solid state nuclear spins \cite{OSullivan2024} or superconducting qubits \cite{Gouzien2021}. Furthermore, rare-earth ion systems have emerged as a compelling platform for optomechanics ~\cite{Molmer2016,Seidelin2019,Chauvet2023} and have demonstrated promise as an alternative to ultrastable cavities in the realm of optical frequency metrology~\cite{Julsgaard2007,Thorpe2011,Galland2020_OL,Lin2023_OE,Lin2024_PRL}.
Here, we will highlight their applications in optical frequency metrology, though our findings also are of general importance across much broader areas of their use.

Frequency stabilization relies on the fact that by selectively burning a spectral hole in the inhomogeneous absorption profile of these ions, an ultra-narrow and temporally stable spectral feature can be created. Despite involving $10^{13}$ ions or more, this spectral hole effectively mimics the homogeneous linewidth of a single ion, providing a strong signal while retaining excellent coherence. This stable spectral structure can then serve as a reliable reference for locking a laser, enabling the generation of an ultra-stable optical frequency reference — particularly valuable for probing optical atomic clocks, which demand high short-term stability, typically over 1 second. The performance of such frequency references depend critically on the stability of the spectral hole over time, not only in terms of its central frequency but also its linewidth, both of which are key to robust frequency locking. One major challenge is mitigating shifts in the central frequency caused by temperature fluctuations. These fluctuations induce a significant frequency shift, proportional to $T^4$, due to two-phonon Raman processes~\cite{Konz2003}. We have previously demonstrated the use of a helium buffer gas to minimize temperature-induced frequency drifts in the 3–4 K range~\cite{Zhang2023_PRA}. Another approach is to further lower the temperature to the sub-kelvin range, using a dilution refrigerator. In our recent work, we showed that this approach effectively renders temperature fluctuations a non-limiting factor for frequency stability in all practical applications~\cite{Lin2024_PRL}.

In this article, we investigate the behavior of spectral hole shapes, a critical factor for enhancing frequency locking performance, at sub-1-kelvin temperatures. This ultra-low temperature regime represents a promising new frontier for applications requiring exceptional frequency stability. More specifically, the optical phase versus frequency dispersion curve corresponding to the spectral hole's absorption profile can be used to produce an error signal suitable for frequency locking. The narrower the spectral hole, for a given contrast, the steeper the slope of this dispersion signal at its center frequency. As the efficiency of detection noise rejection is proportional to this slope~\cite{Pound1983}, this parameter warrants special attention. Furthermore, we investigate how the spectral hole shape is directly influenced by temperature in this regime. These measurements not only aid in evaluating the impact of temperature on frequency stabilization but also offer valuable insights into the crystal structure. Such insights could prove beneficial for future crystal fabrication processes aimed at creating ultra-narrow linewidth emitters.

We employ a system consisting of $\rm Eu^{3+}$ ions doped into a $\rm Y_2SiO_5$ host matrix (Eu:YSO), using the optical transition $^7F_0 \rightarrow$  $^5D_0$. This system offers a linewidth potentially down to 122\,Hz at 1.4\,K, as deduced from photon-echo measurements performed with 100\,G bias magnetic field~\cite{Equall1994}. The $\rm Eu^{3+}$ ions can substitute the $\rm Y^{3+}$ ions in two different crystallographic sites, referred to as site 1 and 2, with vacuum wavelengths of 580.04\,nm and 580.21\,nm, respectively. As the site 1 holds promise for better frequency stability~\cite{Lin2024_PRL}, we will in the following concentrate on this site alone. Our crystal, made by a Czochralski process followed by annealing, contains a 0.1\,at.\% europium doping concentration, and has the dimensions $8\times8\times4\!$\,mm$^3$ and is probed perpendicular to the square facet and parallel to the crystallographic $b$ axis. Spectral holes are created within the $\sim 2$\,GHz inhomogeneous line by resonantly optically pumping $\rm Eu^{3+}$ ions from the $^7F_0$ to the $^5D_0$ state. The ions then decay radiatively to the $^7F_1$ state~\cite{Konz2003}, followed by non-radiative decay into the three hyperfine states of the $^7F_0$ manifold. Optical pumping prevents population accumulation in the particular hyperfine level resonant with the pump beam, generating a transparent window at this specific frequency within the inhomogeneous profile. By scanning the probe laser near the pump-laser frequency, the shape and frequency of the spectral hole are recorded.

\begin{figure*}[t]
\centering
\includegraphics[width=160mm]{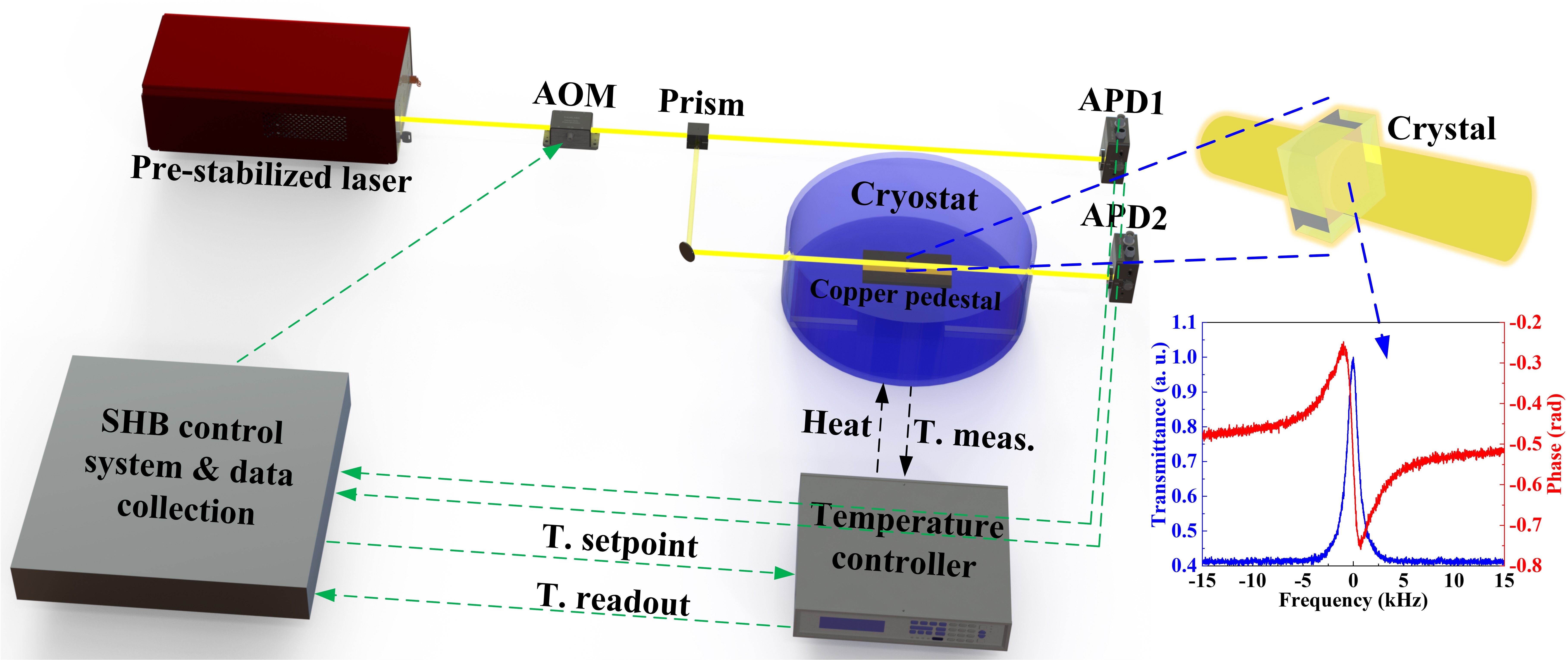}
\caption{\label{exp} Schematics of the main experimental elements for the study of spectral holes in Eu:YSO at dilution refrigerator temperatures specifically used in this work. The laser system, which includes frequency pre-stabilization and frequency doubling (not shown), is indicated in the top left. The setup for optical probing using two avalanche photodiodes (APD), with a zoom on the crystal, is shown on the top right, as well as examples of recorded spectral holes and corresponding dispersion profiles (bottom right). Temperature setpoint and temperature readout are also indicated (temperature abbreviated by T).}%
\end{figure*}

The main elements specific to the work described here are shown in Fig. \ref{exp}. More detailed information concerning our basic experimental setup, including laser-, cryogenics- and control system and can be found in former publications~\cite{Gobron2017,Galland2020_OL,Lin2023_OE}. In brief, an extended cavity diode laser at 1160\,nm is prestabilized, with a controllable offset, onto a room-temperature Fabry-Perot cavity, achieving a fractional frequency instability of $\mathrm{\sim2\times10^{-15}}$ at 1\,s. After frequency doubling to reach the resonances of the Eu$^{3+}$ ions, the laser beam is further manipulated with an acousto-optic modulator for spectroscopy. Differential measurements, with and without the crystal, are carried out by collecting data from avalanche photodiodes placed in the respective paths. A software-defined-radio-based frequency control and data collection system~\cite{Lin2023_OE} steers the offset frequency and amplitude of the laser and generates the desired frequency modes for spectroscopy. In this work, to reach temperatures below 3.5\,K, we use a dilution refrigerator stage inserted into our pulse-tube-based system, allowing for temperatures in the 100\,mK to 1\,K range~\cite{Lin2024_PRL}. 

\begin{figure*}[t]
\centering 
\includegraphics[width=160mm]{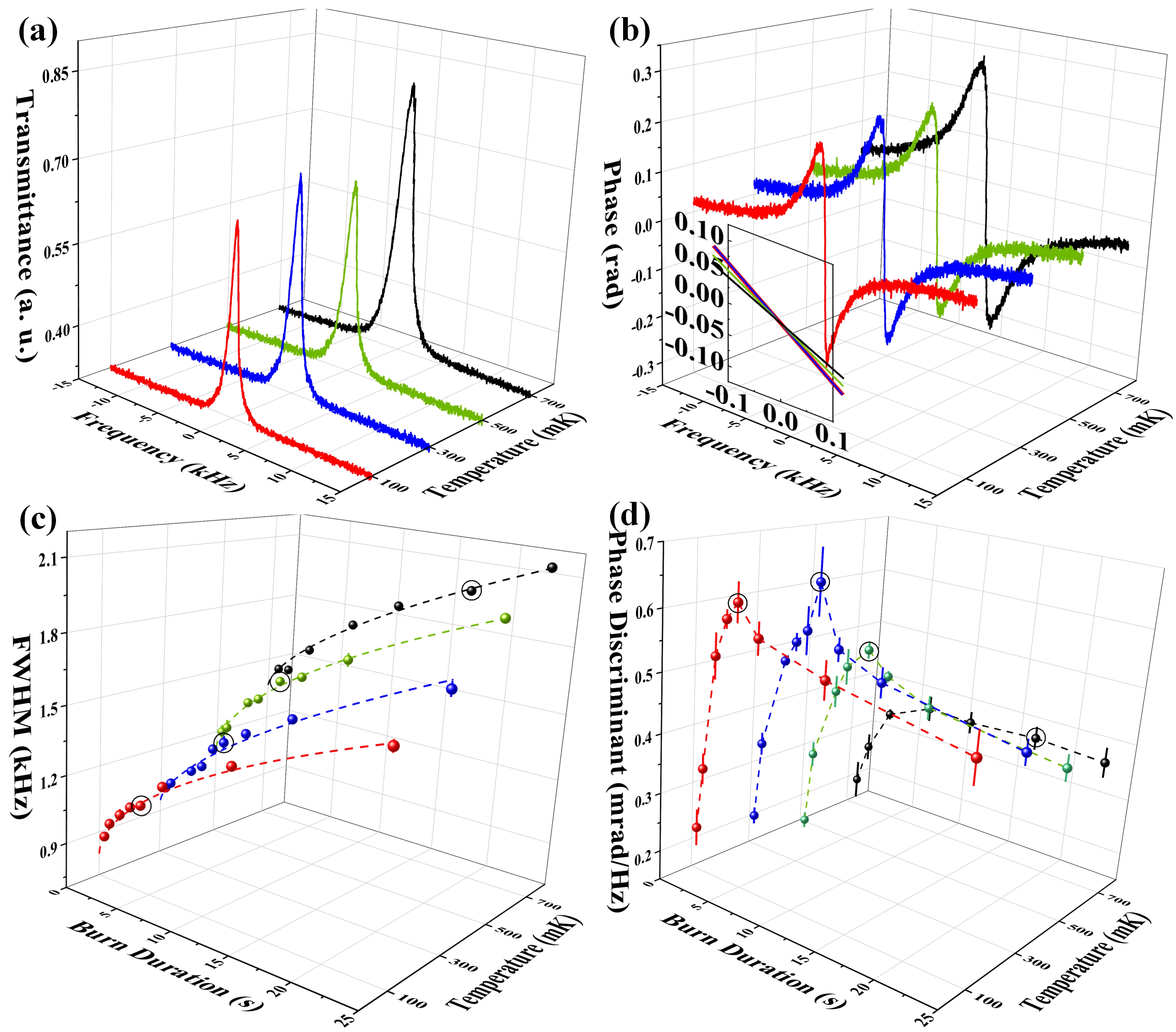}
\caption{\label{width_discrim} Top: Measured spectral hole absorption profiles (a) and phase dispersion curves (b) at different temperatures, with the inset corresponding to a zoom on linear fits of the central part of the dispersion curves (centered in frequency and all temperatures superposed). Bottom: Measurements of spectral hole linewidths (c) and corresponding phase discriminants (d) for different temperatures, as a function of hole burn duration. The black circles around data points correspond to the hole profiles and phase dispersion curves plotted in (a) and (b) above.  The dashed lines in (c) correspond to ad hoc fits ($ax^b+x_0$), which allows us to extrapolate the linewidth to zero burn duration. The dashed lines in (d) connect the data points and are only for guiding the eye. All holes have been burned with 100\,nW laser power, and all data corresponds to the crystallographic site 1.}
\end{figure*}

Operating at a few hundred millikelvin or lower not only reduces temperature-induced frequency fluctuations~\cite{Lin2024_PRL}, but also sharpens the definition of the spectral holes. As mentioned above, this enhancement is especially beneficial for reducing the impact of detection noise, thereby contributing to improving frequency stability. We therefore focus on determining the critical parameters in this context, i.e. the spectral hole linewidth and, more importantly, the center slope of the associated dispersion signal, which quantifies its capability to discriminate small frequency detunings, essential for effective frequency locking. This quantity, which we will refer to as the phase discriminant, is defined as the phase shift that an optical probe undergoes when it is tuned near the center of a narrow spectral hole, as a function of the small detuning from the exact position of the center, as defined by the phase-shift zero-crossing optical frequency. Note that the spectral hole linewidth alone is therefore not the most relevant performance metric: the actual key parameter for frequency locking is the phase discriminant, of which the linewidth is only a component. For example, a very narrow but shallow spectral hole performs less effectively than a slightly broader yet much deeper one. The increased depth of the latter enables a more precise determination of its center, as quantitatively expressed by the phase discriminant. Typical examples of spectral holes and discriminants are presented in Fig. \ref{width_discrim} (a) and (b).

First, to reach the optimal phase discriminants at different temperatures, we determine the range of suitable parameters (intensity and duration) for the spectral hole burning sequence, as these can influence the outcome through instantaneous spectral diffusion in the Eu:YSO crystal \cite{Huang1989,Konz2003} which in turn can broaden the spectral hole, and consequently also reduce the phase discriminant. We find that for optical powers delivered to the crystal of 100\,nW (and below), we observe no effects of spectral hole broadening. This value corresponds to a Gaussian beam of about 3\,mm waist ($1/e^2$ radius) and peak intensity of $\sim$ 710\,nW/cm$^2$. The results are shown in Fig. \ref{width_discrim} where we plot the spectral hole widths and phase dispersion signals measured for different temperatures as a function of burn duration (c and d), and examples of the corresponding hole transmittances and phase dispersion signals from which these values are extracted (a and b). Note that all burn durations are sufficiently long to prevent any additional broadening potentially caused by the Fourier limit. As expected, the spectral hole widths decrease with decreasing temperature, whereas the phase discriminant increases. The optimal burn duration is found in the interval from 4 to 6 seconds, within which phase discriminant reaches a maximum. For temperatures in the 100\,mK to 300\,mK range, we were able to obtain phase discriminant values as high as (0.64 $\pm$ 0.06)\,mrad/Hz, which is more than a two-fold improvement compared to our previously reported work at 3.5\,K \cite{Galland2020_OL}. This can immediately decrease the impact of detection noise in particular, by a factor of two or more compared to our previous assessment~\cite{Lin2023_OE}. We emphasize that this result is obtained without the use of strong polarizing magnetic fields, which are commonly utilized in other work \cite{Oswald2018} and which could also improve our result even further. It is also interesting to note that such a high phase discriminant corresponds to a group delay of approximately 100\,$\mu$s, i.e. a speed of the light going through the 4\,mm long crystal as slow as 40 m/s, which is comparable to complex, specifically engineered systems such as those based on electromagnetically induced transparency \cite{Boller1991,Hau1999}.\\

\begin{figure*}[t] 
\centering
\includegraphics[width=160mm]{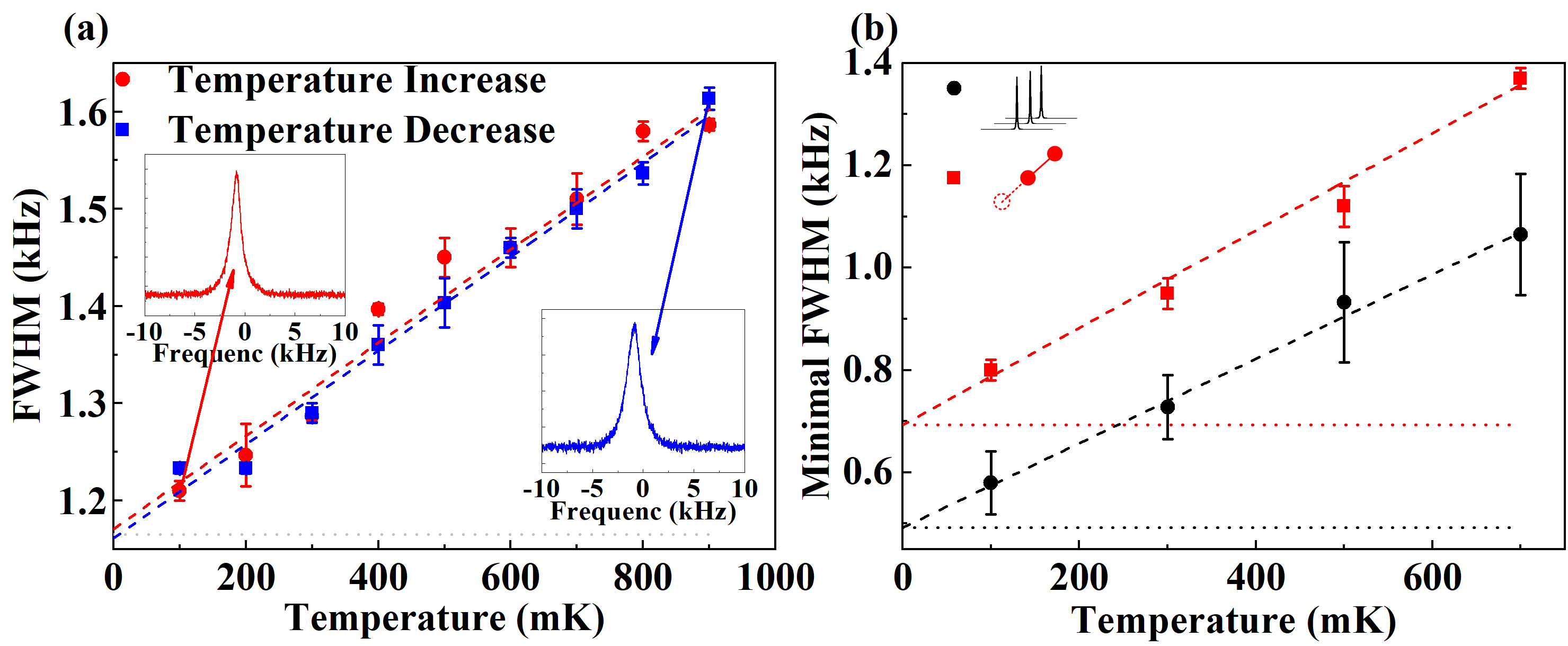}
\caption{\label{width_T} Measurements of the Full Width at Half Maximum (FWHM) of spectral holes as a function of temperature with linear fits. In (a), a single hole was burned (using 100 nW laser power) and then subjected to heating and cooling cycles, while in (b) independent holes for each new temperature are used -- either by adjusting the burn parameters (10 nW laser power) to achieve minimal FWHM for each new temperature (black circles) or by extrapolating the data in Fig.~\ref{width_discrim}(c) (100 nW laser power) to zero burn duration (red squares). Thus, panel (a) tracks the evolution of the same hole, whereas panel (b) ensures that each FWHM measurement is independent of previous thermal history. The dotted (near-horizontal) lines represent the expected linewidth based on two-phonon raman broadening alone. All data correspond to crystallographic site 1.} 
\end{figure*}
 
After establishing the optimal parameters for burning duration and optical power to create the best discriminants at various temperatures, we proceed by studying the linewidth as a function of temperature alone to gain deeper insight into the physical mechanisms governing linewidth behavior. The linewidth of spectral holes in Eu:YSO as a function of temperature has previously been studied above 1.5\,K. We here extend these measurements to the sub 1-kelvin regime, and the results obtained are shown in Fig.~\ref{width_T}. In the first experiment, Fig.~\ref{width_T}(a), a single narrow spectral hole with a discriminant of 0.41\,mrad/Hz was burned at 100\,mK, then progressively heated in 100\,mK steps up to 1\,K, followed by cooling in the same increments. As the data demonstrate, the hole’s linewidth returns to its initial value upon cooling, indicating that no significant temperature-independent broadening occurs, such as broadening caused by repeated probing, which potentially could result in overburning of the spectral hole as a cumulative effect. Note that in this first experiment, we have probed the same spectral hole, burned at 100 mK, across different temperatures. Therefore, these linewidths do not correspond to holes which are burned at the different temperatures indicated in Fig.~\ref{width_T}(a). As there exists a possibility that the dynamics of the hole-burning process may depend on the instantaneous temperature at the moment of burning, the temperature could also indirectly influence the linewidth due to this effect. Thus, in order to assess the linewidth behaviour corresponding to holes burned at the given temperatures as well, we conducted a second experiment in which we burned independent holes at each temperature and immediately probed them at that same temperature. This approach ensures that the linewidth reflects the crystal's temperature at the moment the hole is burned. Additionally, this method eliminates potential contributions strain-induced linewidth modifications \cite{Galland2020, Zhang2020} that could result from temperature changes. As an example, we show in Fig.~\ref{width_T}(b), black circles, the case of holes burned independently at different temperatures, where, in each case, we created the hole with the minimal linewidth possible which still maintains a sufficient signal-to-noise ratio for reliable detection. This was done by adjusting the burn duration to  0.1\,s and the power to 10\,nW. Finally, as an alternative approach, we also extrapolated the fits in Fig.~\ref{width_discrim}(c) to zero burn duration in order to determine a minimum linewidth for different temperatures. The result is shown in Fig.~\ref{width_T}(b), red squares.

For all curves, we observe a clear linear dependence of linewidth as a function of temperature. According to theory and verified experimentally at higher temperatures \cite{Konz2003,Oswald2018}, the two-phonon Raman broadened linewidth of a spectral hole follows the $\Gamma_{0} + \alpha \times T^7$ relation, where $\Gamma_{0}$ is the zero-temperature linewidth, and $\alpha$ was measured by other groups to be 0.0072\,Hz/K$^7$  \cite{Konz2003} or 0.044\,Hz/K$^7$ \cite{Oswald2018} (according to Ref. \cite{Oswald2018}, the difference between these coefficients is possibly due to the varying measurement techniques). In our case, even using the highest reported $\alpha$ value, the linewidth increase from the two-phonon Raman from 0 to 1\,K amounts to just 0.044\,Hz, which is not only negligible compared to the zero temperature linewidth, but also negligible compared to the increase we observe (see dotted lines in Fig.~\ref{width_T}). Thus, no significant two-phonon Raman induced broadening is expected to be visible within this temperature range. Instead, we observe in all three cases a linear increase in the linewidth with temperature, although the prefactors differ significantly. In Fig.~\ref{width_T}(a), the coefficient obtained is (0.48 $\pm$ 0.04)\,kHz/K, while in Fig.~\ref{width_T}(b), we obtain (0.82 $\pm$  0.05)\,kHz/K, and (0.95 $\pm$ 0.06)\,kHz/K for the black circles and red squares, respectively. Note that we do not expect a quantitative agreement among the three obtained coefficients, as the procedures differ significantly. Nevertheless, the shared linear behavior combined with coefficients of the same order of magnitude strongly suggests an underlying mechanism that warrants further investigation. 

Interestingly, a linear behavior, observed in photon-echo experiments, has previously been reported in europium~\cite{Flinn1993,Macfarlane2004,Kunkel2016} and other rare-earth-doped crystals~\cite{Macfarlane2000} at higher temperatures (1.5 to 5.5 K), and below 1 K in silicon-vacancy centers in diamond~\cite{Becker2017}. The authors of these studies primarily attribute the linear dependence to disorder modes or two-level systems (TLS), generally only observed in amorphous materials. These TLS give rise to a broad spectrum of excitation frequencies, with an approximately constant density of states~\cite{Flinn1993}. In reference~\cite{Macfarlane2004}, the authors observed this linear dependence only in some of their Eu:YSO samples, which they referred to as anomalous crystals, based on their general properties. However, based on the properties of our crystal, such as its inhomogeneous linewidth, it seems unlikely that our sample qualifies as anomalous based on this definition, and yet we still observe a linear dependence. This suggests that the presence of TLS may be a matter of degree, dependant on exact crystal fabrication process, rather than an absolute all or nothing. Further complementary investigations are needed to fully understand the physical mechanisms responsible for the linewidth variation at these very low temperatures. For instance, a highly resolved measurement of the crystal's specific heat capacity as a function of temperature between 100\,mK and 1\,K should be able to reveal the presence of TLS \cite{Enss2005}, if such features exist. 

In this article, we investigated spectral holes in $\rm Eu^{3+}:Y_2SiO_5$ at sub-1 kelvin temperatures. By carefully optimizing the burning duration and power, we demonstrated the ability to produce ultranarrow holes with high contrast, enabling phase discriminants that are more than twice as sensitive as those previously observed at 3.5\,K. The best value achieved (0.64\,mrad/Hz) corresponds to an effective light speed through the crystal as low as 40 m/s. Additionally, we examined the temperature dependence of the linewidth and found a linear relationship in this low-temperature regime. This finding, although requiring further investigation, may contribute to a deeper understanding of thermodynamic effects at the microscopic level, and could be of benefit for improving further the crystals in the future.

\vspace{0.2 cm}
 
 S.S. thanks Christophe Marcenat and Thierry Klein for fruitful discussions. The authours acknowledge financial support from Ville de Paris Emergence Program, the Région Ile de France DIM C’nano and SIRTEQ, the LABEX Cluster of Excellence FIRST-TF (ANR-10-LABX-48-01) within the Program ``Investissement d’Avenir'' operated by the French National Research Agency (ANR), the 15SIB03 OC18 and 20FUN08 NEXTLASERS projects from the EMPIR program co-financed by the Participating States and from the European Union’s Horizon 2020 research and innovation program, and the UltraStabLaserViaSHB (GAP-101068547) from the Marie Skłodowska Curie Actions (HORIZON-TMA-MSCA-PF-EF) from the European Commission Horizon Europe Framework Programme (HORIZON).


\begin{thebibliography}{99}

\bibitem{Yano1991} Ryuzi Yano, Masaharu Mitsunaga and Naoshi Uesugi, Ultralong optical dephasing time in Eu$^{3+}$:Y$_{2}$SiO$_{5}$, Optics Letters {\bf 16}, 1884 (1991).

\bibitem{Equall1994} R. W. Equall, Y. Sun, R. L. Cone, and R. M. Macfarlane, Ultraslow optical dephasing in Eu$^{3+}$:Y$_{2}$SiO$_{5}$, Phys. Rev. Lett. {\bf 72}, 2179 (1994).

\bibitem{Zhong2017} T. Zhong, J.M. Kindem, J.G. Bartholomew, J. Rochman, I. Craiciu, E. Miyazono, M. Bettinelli, E. Cavalli, V. Verma, S. W. Nam, F. Marsili, M. D. Shaw, A. D. Beyer, and A. Faraon, Nanophotonic rare-earth quantum memory with optically controlled retrieval, Science {\bf 357} (6358), 1392 (2017).

\bibitem{Laplane2017} C. Laplane, P. Jobez, J. Etesse, N. Gisin, and M. Afzelius, Multimode and Long-Lived Quantum Correlations Between Photons and Spins in a Crystal, Phys. Rev. Lett. {\bf 118}, 210501 (2017).

\bibitem{Chao2020} Chao Liu, Tian-Xiang Zhu, Ming-Xu Su, You-Zhi Ma, Zong-Quan Zhou, Chuan-Feng Li, and Guang-Can Guo, On-Demand Quantum Storage of Photonic Qubits in an On-Chip Waveguide, Phys. Rev. Lett. {\bf 125}, 260504 (2020).

\bibitem{Rakonjac2021} Entanglement between a Telecom Photon and an On-Demand Multimode Solid-State Quantum Memory, Jelena V. Rakonjac, Dario Lago-Rivera, Alessandro Seri, Margherita Mazzera, Samuele Grandi, and Hugues de Riedmatten, Phys. Rev. Lett. {\bf 127}, 210502 (2021).

\bibitem{Rivera2023} Dario Lago-Rivera, Jelena V. Rakonjac, Samuele Grandi and Hugues de Riedmatten, Long distance multiplexed quantum teleportation from a telecom photon to a solid-state qubit, Nature Communications {\bf 14}, 1889 (2023).

\bibitem{OSullivan2024} James O'Sullivan, Jaime Travesedo, Louis Pallegoix, Zhiyuan W. Huang, Alexande May, Boris Yavkin, Patrick Hogan, Sen Lin, Renbao Liu, Thierry Chaneliere, Sylvain Bertaina, Philippe Goldner, Daniel Esteve, Denis Vion, Patrick Abgrall, Patrice Bertet and Emmanuel Flurin, Individual solid-state nuclear spin qubits with coherence exceeding seconds, https://doi.org/10.48550/arXiv.2410.10432

\bibitem{Gouzien2021} Élie Gouzien and Nicolas, Factoring 2048-bit RSA Integers in 177 Days with 13436 Qubits and a Multimode Memory, Phys. Rev. Lett. {\bf} 127, 140503 (2021).

\bibitem{Molmer2016} K. M\o lmer, Y. Le Coq, and S. Seidelin, Dispersive coupling between light and a rare-earth-ion-doped mechanical resonator, Phys. Rev. A  {\bf 94}, 053804 (2016)

\bibitem{Seidelin2019}  S. Seidelin, Y. Le Coq and K. M{\o}lmer, Rapid cooling of a strain-coupled oscillator by an optical phase-shift measurement, Phys. Rev. A {\bf 100}, 013828 (2019)

\bibitem{Chauvet2023} Piezo-orbital backaction force in a rare-earth-doped crystal, A. Louchet-Chauvet, P. J.-P. Poizat, and T. Chanelière, Phys. Rev. Applied {\bf 20}, 054004 (2023).

\bibitem{Julsgaard2007} B. Julsgaard, A. Walther, S. Kr\" oll, and L. Rippe, Understanding laser stabilization using spectral hole burning, Optics Express {\bf 15}, 11444 (2007). 

\bibitem{Thorpe2011} M. J. Thorpe, L. Rippe, T. M. Fortier, M. S. Kirchner, and T. Rosenband, Frequency stabilization to $6\times 10^{-16}$ via spectral-hole burning, Nature Photonics {\bf 5}, 688 (2011).

\bibitem{Galland2020_OL} N. Galland, N. Lu{\v c}i\'c, S. Zhang, H. Alvarez-Martinez, R. Le Targat, A. Ferrier, P. Goldner, B. Fang, S. Seidelin and Y. Le Coq, Double-heterodyne probing for ultra-stable laser based spectral hole burning in a rare-earth doped crystals, Optics Letters {\bf 45}, 1930 (2020).

\bibitem{Lin2023_OE} X. Lin, M. T. Hartman, S. Zhang, S. Seidelin, B. Fang, and Y. Le Coq,  Multi-mode heterodyne laser interferometry realized via software defined radio, Optics Express {\bf 31}, 38475-38493 (2023).

\bibitem{Lin2024_PRL} X. Lin, M. T. Hartman, B. Pointard, R. Le Targat, P. Goldner, S. Seidelin, B. Fang, and Y. Le Coq, Anomalous subkelvin thermal frequency shifts of ultranarrow linewidth solid state emitters, Phys. Rev. Lett. {\bf 133}, 183803 (2024).

\bibitem{Konz2003} F. K\"onz, Y. Sun, C. W. Thiel, R. L. Cone, R. W. Equall, R. L. Hutcheson, and R. M. Macfarlane, Temperature and concentration dependence of optical dephasing, spectral-hole lifetime, and anisotropic absorption in Eu$^{3+}$:Y$_{2}$SiO$_{5}$, Phys. Rev. B  {\bf 68}, 085109 (2003).

\bibitem{Zhang2023_PRA} S. Zhang, S. Seidelin, R. Le Targat, P. Goldner, B. Fang, and Y. Le Coq, First-order thermal insensitivity of the frequency of a narrow spectral hole in a crystal, Phys. Rev. A {\bf 107}, 013518 (2023).

\bibitem{Pound1983} R. V Pound and  R. H. Drever,  A resonant method for measuring the frequencies of light near an atomic resonance. Applied Optics  {\bf 22}, 1965-1968 (1983).

\bibitem{Gobron2017} O. Gobron, K. Jung, N. Galland, K. Predehl, R. Letargat, A. Ferrier, P. Goldner, S. Seidelin and Y. Le Coq, Dispersive heterodyne probing method for laser frequency stabilization based on spectral hole burning in rare-earth doped crystals, Optics Express  {\bf 25}, 15539 (2017).

\bibitem{Huang1989} Jin Huang, J. M. Zhang, A. Lezama, and T. W. Mossberg, Excess dephasing in photon-echo experiments arising from excitation-induced electronic level shifts, Phys. Rev. Lett. {\bf 63}, 78 (1989).

\bibitem{Oswald2018} René Oswald, Michael G. Hansen, Eugen Wiens, Alexander Yu. Nevsky and Stephan Schiller, Characteristics of long-lived persistent spectral holes in Eu$^{3+}$:Y$_{2}$SiO$_{5}$ at 1.2 K, Physical Review A {\bf 98}, 062516 (2018).

\bibitem{Boller1991} K.-J. Boller, A. Imamoğlu, and S. E. Harris, Observation of electromagnetically induced transparency, Phys. Rev. Lett. {\bf 66}, 2593 (1991).

\bibitem{Hau1999} Lene Vestergaard Hau, S. E. Harris, Zachary Dutton and Cyrus H. Behroozi, Light speed reduction to 17 metres per second in an ultracold atomic gas, Nature, {\bf 397}, 594-598 (1999). 

\bibitem{Galland2020} N. Galland, N. Lučić, B. Fang, S. Zhang, R. Letargat, A. Ferrier, P. Goldner, S. Seidelin and Y. Le Coq,  Mechanical tunability of an ultranarrow spectral feature via uniaxial stress, Phys. Rev. Applied {\bf 13}, 044022 (2020).

\bibitem{Zhang2020} S. Zhang, N. Galland, N. Lučić, R. Letargat, A. Ferrier, P. Goldner, B. Fang, Y. Le Coq and S. Seidelin, Inhomogeneous response of an ion ensemble from mechanical stress, Phys. Rev. Research {\bf 2}, 013306 (2020).

\bibitem{Flinn1993} G. P. Flinn, K. W. Jang, Joseph Ganem, M. L. Jones, R. S. Meltzer and R. M. Macfarlane, Sample-dependent optical dephasing in bulk crystalline samples of Y$_{2}$O$_{3}$:Eu$^{3+}$, Phys. Rev. B {\bf 49}, 5821 (1994).

\bibitem{Macfarlane2004} R.M. Macfarlane, Y. Sun, R.L. Cone, C.W. Thiel and R.W. Equall, Optical dephasing by disorder modes in yttrium orthosilicate (Y$_{2}$SiO$_{5}$) doped with Eu$^{3+}$, Journal of Luminescence  {\bf 107}, 310-313 (2004).

\bibitem{Kunkel2016} Nathalie Kunkel, John Bartholomew, Sacha Welinski, Alban Ferrier, Akio Ikesue, and Philippe Goldner, Dephasing mechanisms of optical transitions in rare-earth-doped transparent ceramics, Phys. Rev. B {\bf 94}, 184301 (2016).

\bibitem{Macfarlane2000} R.M. Macfarlane, Y. Sun, F. Könz and R.L. Cone,  Spectral hole burning and optical dephasing in disordered crystals, Pr$^{3+}$:LiNbO$_3$ and Pr$^{3+}$:Sr$_{.6}$Ba$_{.4}$Nb$_2$O$_6$ (SBN),  Journal of Luminescence {\bf 86}, 311-315 (2000). 

\bibitem{Becker2017} Jonas N. Becker, Benjamin Pingault, David Groß, Mustafa Gündoğan, Nadezhda Kukharchyk, Matthew Markham, Andrew Edmonds, Mete Atatüre, Pavel Bushev, and Christoph Becher, All-Optical Control of the Silicon-Vacancy Spin in Diamond at Millikelvin Temperatures, Phys. Rev. Lett. {\bf 120}, 053603 (2018).

\bibitem{Enss2005} C. Enss and S. Hunklinger, Low-Temperature Physics, Springer, 2005.

\end{thebibliography}
\end{document}